\documentclass[aps,prb,twocolumn,showpacs]{revtex4}

\begin{document}
\title{Natural optical activity of metals}

\author{V. P. Mineev$^1$ and Yu Yoshioka$^2$}

\affiliation{$^{1}$ Commissariat \`a l'Energie Atomique,
INAC/SPSMS, 38054 Grenoble, France\\ $^{2}$ Osaka University, Graduate School of Engineering Science, Toyonaka, Osaka 560-8531, Japan}
\date{\today}

\begin{abstract}
We derive the current response of 
noncentrosymmetric metal in normal and in superconducting states to the electromagnetic field with finite frequency and wave vector.  The conductivity tensor is found contains the linear in wave vector off diagonal part generating natural optical activity. 
The 
Kerr rotation of polarization of light reflected from the metal surface is calculated.
Its value is expressed through the fine-structure constant and the ratio of light frequency to the band splitting due to the spin-orbit interaction.

\end{abstract}


\maketitle

The natural optical activity or natural gyrotropy is well known phenomenon typical for the bodies having no centre of symmetry \cite{LL}. In this case, the tensor of dielectric permeability  has  linear terms in the expansion in powers of wave vector
\begin{equation}
\varepsilon_{ij}(\omega,{\bf q})=\varepsilon_{ij}(\omega,0)+i\gamma_{ijl}q_l,
\label{e1}
\end{equation}
where $\gamma_{ikl}$ is an antisymmetric third rank tensor called the tensor of gyrotropy. The optical properties of a naturally active 
body resemble those of the magneto-active media having no time reversal symmetry. It exhibits double circular refraction, the Faraday and the Kerr effects. 

The description of the natural optical activity
in terms of linear spacial dispersion of permeability \cite{LL} is appropriate for solid or liquid dielectric media. Whereas in the case of metals, it is more natural to formulate them in terms of spacial dispersion of conductivity tensor having the following form:
\begin{equation}
\sigma_{ij}(\omega,{\bf q})=\sigma_{ij}(\omega,0)-i\lambda_{ijl}q_l.
\label{e2}
\end{equation}

The metals without inversion symmetry have recently become a subject of considerable interest 
arising mostly due to the discovery of superconductivity in CePt$_3$Si.
\cite{Bauer04}
Now the list of noncentrosymmetric superconductors has grown to
include UIr \cite{Akazawa04}, CeRhSi$_3$ 
\cite{Kimura05}, CeIrSi$_3$ \cite{Sugitani06},
Y$_2$C$_3$ \cite{Amano04},
Li$_2$(Pd$_{1-x}$,Pt$_x$)$_3$B \cite{LiPt-PdB},
KOs$_2$O$_6$ \cite{KOsO}, and other compounds. The theory has been mostly developed
for description of superconducting properties of such a type of materials (for  review see Ref. \onlinecite{MinSig}).

The spin-orbit coupling of electrons in noncentrosymmetric crystal lifts spin the degeneracy of the electron energy band causing a noticeable band splitting. The Fermi surface splitting can be observed by the de Haas-van Alphen effect discussed theoretically in the paper Ref. \onlinecite{MinSam05}.  The band splitting reveals itself in the large residual value of the spin susceptibility of noncentrosymmetric superconductors at zero temperature.\cite{Sam07} It also makes possible the existence of nonuniform superconducting states those can be traced to the Lifshitz invariants in the free energy.\cite{MinSam08}
 Another significant
manifestation of the band splitting is the natural optical activity.
Here, we present the derivation of general expression for the current response to the electro-magnetic field with finite frequency and wave vector valid for the normal and the superconducting state of noncentrosymmetric metals. We apply this result to calculation of   
Kerr rotation of polarization of light reflected from the surface of metal with cubic symmetry.
In this case, the usual part of the conductivity tensor is isotropic 
$\sigma_{ij}(\omega,0)=\sigma(\omega)\delta_{ij}$ and the gyrotropic conductivity tensor $\lambda_{ikl}=\lambda e_{ikl}$ is determined by the single complex coefficient  
 $\lambda=\lambda'+i\lambda''$ such that a 
normal state density of current  is 
\begin{equation}
{\bf j}={\sigma}{\bf E}+\lambda~\text{rot}{\bf E}.
\label{cur}
\end{equation}
We shall find here that the gyrotropy conductivity $\lambda$ is directly proportional to the ratio of the light frequency to the band splitting value.

 The current response to the electromagnetic field at finite 
${\bf q}$ and $\omega$ can be written following the textbook procedure \cite{FizKin}
generalized to two-band case in \cite{FaVa}. In application to our situation,
one has to remember that in a noncentrosymmetric crystal, all the values like
single electron energy 
\begin{equation}
\xi_{\alpha\beta}({\bf k})=\xi_0({\bf k})\delta_{\alpha\beta}+
\mbox{\boldmath$\gamma$}({\bf k})\mbox{\boldmath$\sigma$}_{\alpha\beta},
\end{equation}
velocities
${\bf v}_{\alpha\beta}({\bf k})=\partial\xi_{\alpha\beta}({\bf k})/\partial{\bf k}$, the inverse effective mass
$(m^{-1}_{ij})_{\alpha\beta}=\partial^2\xi_{\alpha\beta}({\bf k})/\partial k_i\partial k_j$, the Green functions 
$G_{\alpha\beta}(\tau_1,{\bf k};\tau_2,{\bf k}')=-\langle T_\tau a_{{\bf k}\alpha}(\tau_1)a^\dagger_{{\bf k}'\beta}(\tau_2)\rangle$ and $F_{\alpha\beta}(\tau_1,{\bf k};\tau_2,{\bf k}')=\langle T_\tau a_{{\bf k}\alpha}(\tau_1)a_{-{\bf k}'\beta}(\tau_2)\rangle$ are matrices in the spin space (see for instance Ref. \onlinecite{Sam07}). 
Taking this in mind, we obtain
\begin{eqnarray}
\label{basic}
 j_i(\omega_n,{\bf q})=~~~~~~~~~~~~~~~~~~~~~~~~~~~~~~\\
 -e^2Tr\left [\text{T}\sum_{m=-\infty}^{\infty}\int  \frac{d^3k}{(2\pi)^3} 
\right.
\{  \hat v_i({\bf k})\hat G^{(0)}(K_+)\hat v_j({\bf k})\hat G^{(0)}(K_-)\nonumber\\
\left.
+ \hat v_i({\bf k})\hat F^{(0)}(K_+)\hat v_j^t(-{\bf k})\hat F^{+(0)}(K_-)\}
+\hat m^{-1}_{ij}\hat n_e\right ]{A_j}(\omega_n,{\bf q}).\nonumber
\end{eqnarray}
The transposed matrix of velocity is determined as $\hat{\bf v}^t(-{\bf k})=\partial\hat \xi^t(-{\bf k})/\partial{\bf k}$.  The arguments of the zero field Green functions are denoted as $
K_{\pm}=\left(\Omega_m\pm{\omega_n}/{2}, {\bf k}\pm{\bf q}/{2}\right)$. The Matsubara frequencies 
take the values $\Omega_m=\pi (2m+1-n)T$ and $\omega_n=2\pi nT $. We put here $\hbar=c=1$ and return back to the dimensional units in the final expressions.

Unlike to the paper \onlinecite{Min08}, we do not write here the part of the current density  arising due to the gauge shift in the argument of the order parameter $\Delta({\bf k}-e{\bf A}(\omega_n,{\bf q}))$ describing magneto-optical phenomena in the superconductors with broken time-reversal symmetry. We will return to this problem in a separate publication.

Let us pass from the spin to the band representation, where the one-particle Hamiltonian
\begin{equation}
H_0=\sum_{\bf k}\xi_{\alpha\beta}({\bf k})a^{\dagger}_{{\bf k}\alpha}a_{{\bf k}\beta}=
\sum_{{\bf k},\lambda=\pm}\xi_{\lambda}({\bf k})c^{\dagger}_{{\bf k}\lambda}c_{{\bf k}\lambda}
\end{equation}
has diagonal form. Here, the band energies are 
$\xi_{\lambda}({\bf k})=\xi_0({\bf k})+\lambda
|\mbox{\boldmath$\gamma$}({\bf k})|$.
The diagonalization is made by the following transformation
\begin{equation}
\label{band transform}
    a_{{\bf k}\alpha}=\sum_{\lambda=\pm}u_{\alpha\lambda}({\bf k})c_{{\bf k}\lambda},
\end{equation}
with the coefficients
\begin{equation}
\label{Rashba_spinors}
    \begin{array}{l}
    \displaystyle u_{\uparrow\lambda}{(\bf k})=
    \sqrt{\frac{|\mbox{\boldmath$\gamma$}|+\lambda\gamma_z}{2|\mbox{\boldmath$\gamma$}|}},\\
    \displaystyle u_{\downarrow\lambda}({\bf k})=\lambda
    \frac{\gamma_x+i\gamma_y}{\sqrt{2|\mbox{\boldmath$\gamma$}|(|\mbox{\boldmath$\gamma$}|+\lambda\gamma_z)}},
    \end{array}
\end{equation}
forming a unitary matrix $\hat u({\bf k})$.

The zero field Green functions in the band representation are diagonal and have the following form:
\cite{Sam07}
\begin{eqnarray}
G^{(0)}_{\lambda\lambda'}(\omega_n,{\bf k})=\delta_{\lambda\lambda'}G_{\lambda}(\omega_n,{\bf k}),\nonumber\\
F^{(0)}_{\lambda\lambda'}(\omega_n,{\bf k})=\delta_{\lambda\lambda'}F_{\lambda}(\omega_n,{\bf k}),
\end{eqnarray} 
where
\begin{eqnarray}
G_{\lambda}(\omega_n,{\bf k})=-\frac{i\omega_n+\xi_\lambda}{\omega_n^2+\xi_\lambda^2+
|\tilde\Delta_{\lambda}({\bf k})|^2},\nonumber\\
F_{\lambda}(\omega_n,{\bf k})=\frac{t_\lambda({\bf k})\tilde\Delta_{\lambda}({\bf k})}{\omega_n^2+\xi_\lambda^2+
|\tilde\Delta_{\lambda}({\bf k})|^2},
\end{eqnarray} 
and
\begin{equation}
\label{t}
    t_{\lambda}({\bf k})=-\lambda
    \frac{\gamma_x({\bf k})-i\gamma_y({\bf k})}{\sqrt{\gamma_x^2({\bf k})+\gamma_y^2({\bf k})}}.
\end{equation}
The functions $\tilde\Delta_{\lambda}({\bf k})$ are the gaps in the $\lambda$-band quasiparticle spectrum in superconducting state.
In the simplest model with BCS pairing interaction $v_g({\bf k},{\bf k}')= -V_g$, the gap functions are the same in both bands: $\tilde\Delta_{+}({\bf k})=\tilde\Delta_{-}({\bf k})=\Delta$ and we deal with pure singlet pairing \cite{SamMin08}.

 Transforming the Green functions according the eqn. (\ref{band transform}) to the band representation, we obtain for the trace of matrices in eqn. (\ref{basic}):
 \begin{eqnarray}
 \label{trace}
Tr \{  \hat v_i({\bf k})\hat G^{(0)}(K_+)\hat v_j({\bf k})\hat G^{(0)}(K_-)\nonumber\\
+ \hat v_i({\bf k})\hat F^{(0)}(K_+)\hat v_j^t(-{\bf k})\hat F^{+(0)}(K_-)\}=\nonumber\\
v_{++,i}({\bf k})G_+v_{++,j}({\bf k})G_++v_{++,i}({\bf k})F_+v_{++,j}(-{\bf k})F^\dagger_++\nonumber\\
v_{--,i}({\bf k})G_-v_{--,j}({\bf k})G_-+v_{--,i}({\bf k})F_-v_{--,j}(-{\bf k})F^\dagger_++\nonumber\\
v_{+-,i}({\bf k})G_-v_{-+,j}({\bf k})G_++v_{+-,i}({\bf k})F_-v_{-+,j}(-{\bf k})F^\dagger_++\nonumber\\
v_{-+,i}({\bf k})G_+v_{+-,j}({\bf k})G_-+v_{-+,i}({\bf k})F_+v_{+-,j}(-{\bf k})F^\dagger_-.~~
 \end{eqnarray}
For the brevity, we have omit here the arguments of the Green functions.
They are the same as in the upper two lines.
The matrix velocity in the band representation is 
 \begin{eqnarray}
 {\bf v}_{\lambda\lambda'}(\pm{\bf k})=u^{\dagger}_{\lambda\alpha}(\pm{\bf k}){\bf v}_{\alpha\beta}(\pm{\bf k})u_{\beta\lambda'}(\pm{\bf k})=\nonumber\\
 \frac{\partial \xi_0(\pm{\bf k})}{\partial{\bf k}}\delta_{\lambda\lambda' }+
\frac{\partial\gamma_l(\pm{\bf k})}{\partial{\bf k}}\tau_{l,\lambda\lambda'}(\pm{\bf k}),
 \end{eqnarray}
 where 
$\mbox{\boldmath$\hat \tau$}({\bf k})
=\hat u^{\dagger}({\bf k})\mbox{\boldmath$\hat\sigma$}\hat u({\bf k})$ are hermitian matrices. We neglected the difference between $\hat u({\bf k})$ and $\hat u({\bf k}\pm{\bf q}/2)$, which is of the order of $(k_F\delta)^{-1}$,
where $\delta$ is the skin effect penetration depth.\cite{depth} 

For the calculation of gyrotropy of conductivity, only the last two lines in eqn. (\ref{trace}) consisting of interband terms are important. They are equal to
\begin{eqnarray}
 \label{trace'}
\frac{\partial\gamma_l}{\partial k_i}
\frac{\partial\gamma_m}{\partial k_j}\{\tau_{l,+-}\tau^{*}_{m,+-}
[G_-G_+-F_-F^\dagger_+]+\nonumber\\
\tau_{l,-+}\tau^{*}_{m,-+}
[G_+G_--F_+F^\dagger_-]\}.
 \end{eqnarray}
Using the identity 
\begin{equation}
\tau_{l,+-}\tau^{*}_{m,+-}=\tau^*_{l,-+}\tau_{m,-+}=\delta_{lm}-\hat\gamma_l\hat\gamma_m+ie_{lmn}\hat\gamma_n,
\end{equation}
where $\hat\gamma_l=\gamma_l/|\mbox{\boldmath$\gamma$}|$,
one can rewrite the above expression as
\begin{eqnarray}
 \label{trace1}
\frac{\partial\gamma_l}{\partial k_i}
\frac{\partial\gamma_m}{\partial k_j}\{(\delta_{lm}-\hat\gamma_l\hat\gamma_m)
[G_-G_++G_+G_--F_-F^\dagger_+~~~~~~~\\
-F_+F^\dagger_-]+
ie_{lmn}\hat\gamma_n
[G_-G_+-G_+G_--F_-F^\dagger_++F_+F^\dagger_-]\}.\nonumber
 \end{eqnarray}

The linear part of current with respect to the wave vector ${\bf q}$ originates  from the last term in the eqn. (\ref{trace1}). 
\begin{eqnarray}
 \label{trace2}
j^g_i(\omega_n,{\bf q})=-ie_{ijl}e^2\gamma_0^2\text{T}\sum_{m=-\infty}^{\infty}\int  \frac{d^3k}{(2\pi)^3}\times\nonumber\\ 
\hat \gamma_l[G_-(K_+)G_-(K_-)-G_+(K_+)G_-(K_-)-\nonumber\\ F_-(K_+)F^\dagger_+(K_-)+F_+(K_+)F^\dagger_-(K_-)]{A_j}(\omega_n,{\bf q}).
 \end{eqnarray}
 The momentum dependence of the spin-orbit coupling is determined by the
crystal symmetry. For the cubic group $G=O$,
which describes the point symmetry of
Li$_2$(Pd$_{1-x}$,Pt$_x$)$_3$B, the simplest form compatible with
the symmetry requirements is
\begin{equation}
\label{gamma_O}
 \mbox{\boldmath$\gamma$}({\bf k})=\gamma_0{\bf k},
\end{equation}
where $\gamma_0$ is a constant. 
For the tetragonal group
$G={C}_{4v}$, which is relevant for CePt$_3$Si,
CeRhSi$_3$ and CeIrSi$_3$, the spin-orbit coupling is given by
\begin{equation}
\label{gamma C4v}
\mbox{\boldmath$\gamma$}({\bf k})=\gamma_\perp(k_y\hat x-k_x\hat y)
    +\gamma_\parallel k_xk_yk_z(k_x^2-k_y^2)\hat z.
\end{equation}

One can show that for  the tetragonal group $G={C}_{4v}$,
 the linear  in the component of wave vector ${\bf q}$ part of conductivity is absent.
So, we continue calculation for the metal with cubic symmetry where 
$\hat\gamma={sign}\gamma_0~\hat{k} $.  
In the following, we put  $\hat\gamma=\hat{k} $ taking
$\gamma_0$ as a positive constant.
Let us find first the gyrotropy conductivity in the normal state. Performing summation over the Matsubara frequencies, we obtain
\begin{eqnarray}
j^g_i(\omega_n,{\bf q})=-ie_{ijl}e^2\gamma_0^2
\int  \frac{d^3k}{(2\pi)^3}\hat k_l\times\nonumber\\ 
\left\{\frac{f(\xi_+({\bf k}_-))-f(\xi_-({\bf k}_+))}{i\omega_n+\xi_+({\bf k}_-)-\xi_-({\bf k}_+)}\right.\nonumber\\
\left.-\frac{f(\xi_-({\bf k}_-))-f(\xi_+({\bf k}_+))}{i \omega_n+\xi_-({\bf k}_-)-\xi_+({\bf k}_+)}\right\}
{A_j}(\omega_n,{\bf q}).
 \end{eqnarray}
Here $f(\xi_\pm( {\bf k}_{\pm}))$ is the Fermi distribution function and $ {\bf k}_{\pm}={\bf k}\pm {\bf q}/2$.
By changing  the sign of momentum  ${\bf k}\to-{\bf k} $ in the first term under integral and making use that $\xi_\lambda({\bf k})$ is even function of ${\bf k}$, we come to
\begin{eqnarray}
j^g_i(\omega_n,{\bf q})=-2ie_{ijl}e^2\gamma_0^2
\int  \frac{d^3k}{(2\pi)^3}\hat k_l[\xi_-({\bf k}_-)-\xi_+({\bf k}_+)]\times\nonumber\\ 
\frac{f(\xi_+({\bf k}_+))
-f(\xi_-({\bf k}_-))}
{(\omega_n)^2+(\xi_+({\bf k}_+)-\xi_-(\bf k_-))^2}
{A_j}(\omega_n,{\bf q}).~~
 \end{eqnarray}
Analytical continuation of this expression from the discrete set of Matsubara frequencies into entire 
half-plane $\omega>0$ is performed by the usual substitution $i\omega_n\to\omega+i\delta$ with $\delta\to 0$. 
We see that the gyrotropic part of current is an odd function of the wave vector. At the same time, it is an even function of frequency. Hence, it presents a sort of displacement  current originating of band splitting in noncentrosymmetric metal.

We shall be interested in the low frequency and small $q$ limit of this general formula. Expanding this expression up to the first order in $q_n\omega^2$ and performing integration over momentum space in the limit $\hbar\omega \ll \gamma_0k_F\ll\varepsilon_F$, we obtain 
\begin{equation}
j^g_i(\omega,{\bf q})=-ie_{ijn}\frac{e^2\omega^2 }{12\pi^2\gamma_0k_F}q_n{A_j}(\omega,{\bf q}),
\end{equation}
or, after substitution of the Fourier component of the vector potential by the Fourier component of an electric field
${\bf A}={\bf E}/i\omega$, 
\begin{equation}
j^g_i(\omega,{\bf q})=-e_{ijn}\frac{e^2\omega }{12\pi^2\gamma_0k_F}q_n{E_j}(\omega,{\bf q}).
\end{equation}
This corresponds to
\begin{equation}
\lambda=-i\frac{e^2\omega}{12\pi^2\gamma_0k_F}.
\label{lambda}
\end{equation}

To calculate the Kerr rotation, we consider  linearly polarized light normally incident from vacuum to the boundary of a medium with complex index of refraction
 \begin {equation}
 N=n+i\kappa
 \end{equation}
expressed through the diagonal part of complex conductivity $\sigma=\sigma'+i\sigma''$
by means of the usual relations 
  \begin {equation}
n^2-\kappa^2=1-\frac{4\pi\sigma''}{\omega},~~~~2n\kappa=\frac{4\pi\sigma'}{\omega}.
\label{e2'}
 \end{equation}
The light is reflected as elliptically polarized with the major axis rotated relative to the incident polarization by an amount \cite{Ben}
\begin {equation}
\theta=\frac{(1-n^2+\kappa^2)\Delta\kappa+2n\kappa\Delta n}{(1-n^2+\kappa^2)^2+(2n\kappa)^2},
\label{e3}
\end{equation} 
where for the current given by eqn. (\ref{cur}).
\begin {equation}
\Delta n=n_+-n_-=\frac{4\pi\lambda''}{c},
\label{e4}
\end{equation} 
\begin {equation}
\Delta\kappa=\kappa_+-\kappa_-=\frac{4\pi\lambda'}{c}
\label{e5}
\end{equation} 
are the differences in the real  and imaginary parts of the refraction indices of circularly polarized lights  with the opposite 
polarization. 

After substitution the eqn. (\ref{lambda})   in eqns. (\ref{e5}), (\ref{e4}) 
we find that $\Delta\kappa=0$ and  $\Delta n$
expresses through ratio of the light frequency to the band splitting $2\gamma_0 k_F$ as
\begin {equation}
\Delta n=-\frac{\alpha}{3\pi}\frac{\hbar\omega}{\gamma_0 k_F}.
\label{e6}
\end{equation} 
Here, $\alpha=e^2/\hbar c$ is the fine structure constant.

It is worth to be noted that the coefficient $\lambda$ consists of two parts:
contribution of the quasiparticles in between the Fermi surfaces of two splitting bands
and in the vicinity of the two Fermi surfaces (compare with Ref. \onlinecite{Sam07}). 
In the superconducting state at $\hbar\omega \gg \Delta$, $\lambda$ can be given as
\begin{equation}
        \lambda=-i\frac{e^2\omega}{8\pi^2\gamma_0k_F}\left (1-\frac{1}{3}
                Y(T)\right ),
\end{equation}
and
\begin{equation}
        Y(T)=\int \frac{1}{2T}
        \frac{1}{\cosh^2(\sqrt{\xi^2+\Delta^2}/2T)}d\xi
        \label{lambdasc}
\end{equation}
is the Yosida function.
At $T=0$, the second term in eqn. (\ref{lambdasc}), contribution of the quasiparticles in the vicinity of the Fermi surfaces,
disappears and only the first term remains. As a result, we obtain in superconducting state at $T=0$
\begin{equation}
        \Delta n =-\frac{\alpha}{2\pi} \frac{\hbar \omega}{\gamma_0k_F}.
\end{equation}

In conclusion,  we presented here the derivation of the current response 
to the electromagnetic field with finite frequency and wave vector in noncentrosymmetric metal. The conductivity tensor contains a gyrotropic part responsible for the natural optical activity. As an example the Kerr rotation for the polarized light reflected from the surface of noncentrosymmetric metal with cubic symmetry is calculated.
The found value of the Kerr angle is expressed through the fine structure constant and the ratio of the light frequency to the spin-orbit band splitting. The result can be used for the direct experimental determination of the latter value.
 
One of the authors (V. P. M.) is indebted to E. Kats for the helpful discussion of natural optical activity
and to  L. Falkovsky and A.Varlamov attracted his attention to their paper about the current in  two-band conductors \cite{FaVa}. 
The financial support of  another author (Y. Y.) by the Global COE program (G10) from Japan Society for the Promotion of Science is gratefully acknowledged.


\begin{thebibliography}{99}
\bibitem{depth} The simple calculation taking in mind the current expression eqn. (\ref{cur}) shows that for linearly polarized light, the skin penetration depth is given by usual formula (see the textbook \cite{LL}). However, the skin layer for the circularly polarized light is a bit thicker due to the last gyrotropic term in formula eqn. (\ref{cur}).
\end{thebibliography}
\end{document}